\documentclass[aps,prx,preprint,letterpaper,superscriptaddress,longbibliography, 11pt]{revtex4-2}

\usepackage{amsmath,amsthm,amsfonts}
\usepackage{multirow}
\usepackage{array}
\usepackage{soul}
\usepackage{float}
\usepackage{graphicx}
\usepackage{afterpage}
\usepackage{siunitx}
\usepackage{gensymb}

\usepackage[justification=raggedright,singlelinecheck=false,labelfont=bf,labelsep=quad, format=plain,font=small]{caption,subcaption}

\usepackage[colorlinks,citecolor=blue,urlcolor=blue,bookmarks=false,hypertexnames=true]{hyperref}

\usepackage{cleveref}

\newcommand{\sub}[1]{\textsubscript{#1}}
\newcommand{\degC}[1]{#1 \degree C}
\newcommand{\Ang}{\AA}

\begin{document}

\title{Isotropic plasma-thermal atomic layer etching of superconducting TiN films using sequential exposures of molecular oxygen and SF\sub6/H\sub2 plasma}

\author{Azmain A. Hossain}
\affiliation{Division of Engineering and Applied Science, California Institute of Technology, Pasadena, CA 91125, USA}
\author{Haozhe Wang}
\affiliation{Division of Engineering and Applied Science, California Institute of Technology, Pasadena, CA 91125, USA}
\author{David S. Catherall}
\affiliation{Division of Engineering and Applied Science, California Institute of Technology, Pasadena, CA 91125, USA}
\author{Martin Leung}
\affiliation{Division of Natural Sciences, Pasadena City College, Pasadena, CA 91106, USA}
\author{Harm C. M. Knoops}
\affiliation{Oxford Instruments Plasma Technology, North End, Bristol BS49 4AP, U.K.}
\affiliation{Department of Applied Physics, Eindhoven University of Technology, P.O. Box 513, 5600MB Eindhoven, The Netherlands}
\author{James R. Renzas}
\affiliation{Oxford Instruments Plasma Technology, North End, Bristol BS49 4AP, U.K.}
\author{Austin J. Minnich}
\email{aminnich@caltech.edu}
\affiliation{Division of Engineering and Applied Science, California Institute of Technology, Pasadena, CA 91125, USA}
\date{\today}

\begin{abstract}

Microwave loss in superconducting TiN films is attributed to two-level systems in various interfaces arising in part from oxidation and microfabrication-induced damage.  Atomic layer etching (ALE) is an emerging subtractive fabrication method which is capable of etching with Angstrom-scale etch depth control and potentially less damage. However, while ALE processes for TiN have been reported, they either employ HF vapor, incurring practical complications; or the etch rate lacks the desired control. Further, the superconducting characteristics of the etched films have not been characterized. Here, we report an isotropic plasma-thermal TiN ALE process consisting of sequential exposures to molecular oxygen and an SF\sub{6}/H\sub{2} plasma. For certain ratios of SF\sub{6}:H\sub{2} flow rates, we observe selective etching of TiO\sub{2} over TiN, enabling self-limiting etching within a cycle. Etch rates were measured to vary from 1.1 \Ang/cycle at \degC{150} to 3.2 \Ang/cycle at \degC{350} using ex-situ ellipsometry. We demonstrate that the superconducting critical temperature of the etched film does not decrease beyond that expected from the decrease in film thickness, highlighting the low-damage nature of the process. These findings have relevance for applications of TiN in microwave kinetic inductance detectors and superconducting qubits.

\end{abstract}

\maketitle

\newpage

\section{Introduction}

Titanium nitride (TiN) is a superconducting metal of interest for microelectronics and superconducting quantum devices. Its high kinetic inductance, low microwave loss, and high absorption coefficient in the infrared and optical frequencies make it a promising material for single photon detectors \cite{Leduc2010Sep, Vissers2010Dec}, ultra-sensitive current detectors \cite{Kher2016Jul}, quantum-limited parametric amplifiers \cite{HoEom2012Aug}, and qubits \cite{Hazard2019Jan, Chang2013Jul}. Superconducting microwave resonators based on TiN routinely exhibit internal quality factors Q\sub{i} $> 10^6$  \cite{Chang2013Jul, Shearrow2018Nov, Vissers2010Dec}.  TiN is also used for microelectronic applications in which it is used as a copper diffusion barrier and metal gate electrode \cite{Kim2003Nov, Zhao2019Jun, Lima2012Jun}. In many of these applications, imperfections at film interfaces are the primary limitation to figures of merit for various devices. For instance, the quality factor of superconducting microresonators is presently thought to be limited by microwave surface loss associated with two-level systems (TLS) in various interfaces \cite{Gao2007Mar, Barends2008Jun, Gao2008Apr}. Subtractive nanofabrication methods based on typical wet or dry etching processes are unsuitable for mitigating TLS density in these devices due to the lack of Angstrom-scale precision in etching and the sub-surface damage they induce \cite{Sandberg2012Jun, Altoe2022Apr, Gao2022Mar}.

Atomic layer etching (ALE) is an emerging subtractive nanofabrication process with potential to overcome these limitations \cite{Lill2016, George2020Jun, Sang2020}. Early forms of ALE focused on directional etching \cite{Sakaue1990Nov, Horiike1990May}. Directional ALE is based on surface modification by adsorption of a reactive species, and subsequent sputtering of the modified surface with ions or neutral atoms of low energy exceeding only the sputtering threshold of the modified surface \cite{Kanarik2015, Oehrlein2015Rev3}. Isotropic thermal ALE processes have also been developed recently using sequential, self-limiting surface chemical reactions \cite{George2016}. In thermal ALE, the material surface is modified to form a stable layer that can then be removed by a selective mechanism, such as temperature cycling, ligand-exchange transmetalation reactions, or others \cite{Osakada2003Jan, George2020Jun}. Isotropic thermal and plasma ALE processes have now been reported for various dielectrics and semiconductors including Al\sub{2}O\sub{3} \cite{Zywotko2018Nov, Lee2015May}, SiO\sub{2} \cite{DuMont2017Mar, Rahman2018Sep}, AlN \cite{Johnson2016Sep, Wang2023May, Cano2022Apr}, InGaAs \cite{Lu2019Aug, Ohba2017May} and others \cite{Lee2015Mar, Fang2018Dec, Fischer2021May, George2020Jun}. Surface smoothing of etched surfaces using ALE has also been reported for various metals and semiconductors \cite{Zywotko2018Nov, Kanarik2017Sep, Ohba2017May, Gerritsen2022Dec}.

For TiN, ALE processes based on fluorination and ligand exchange with Sn(acac)\sub{2}, trimethylaluminum (TMA), dimethylaluminum chloride (DMAC), and SiCl\sub{4} did not lead to  etching \cite{Lee2016}. When fluorinated, TiN retains its 3+ oxidation state, yielding TiF\sub{3}. TiF\sub{3} either formed non-volatile ligand-exchange products or did not react with the precursors, and hence no etching occurred. This difficulty was overcome by first converting the Ti to the 4+ oxidation state with exposure to ozone or H\sub{2}O\sub{2}, which upon fluorination using HF produced volatile TiF\sub{4} \cite{Lee2017TiN}. A conceptually similar process has also been reported using O\sub{2} plasma and CF\sub{4} plasma \cite{Shim2022}.  

Despite these advances, limitations remain. The use of HF vapor incurs practical complications. The process of Ref.~\cite{Shim2022} based on O\sub{2} plasma and CF\sub{4} requires a heating and cooling step per cycle which can lead to impractical time per cycle on most conventional plasma tools. Additionally, the recipe achieves nm/cycle etch rates, which lacks the desired Angstrom-scale control and low damage characteristics. Previous reports did not examine the effects of ALE on the superconducting properties of the samples. Identifying alternate reactants to HF vapor while maintaining Angstrom-level precision over the thickness, and ensuring that superconducting properties are not degraded, all remain topics of interest for TiN ALE.

\begin{figure}
    \centering
    \includegraphics[width = \textwidth]{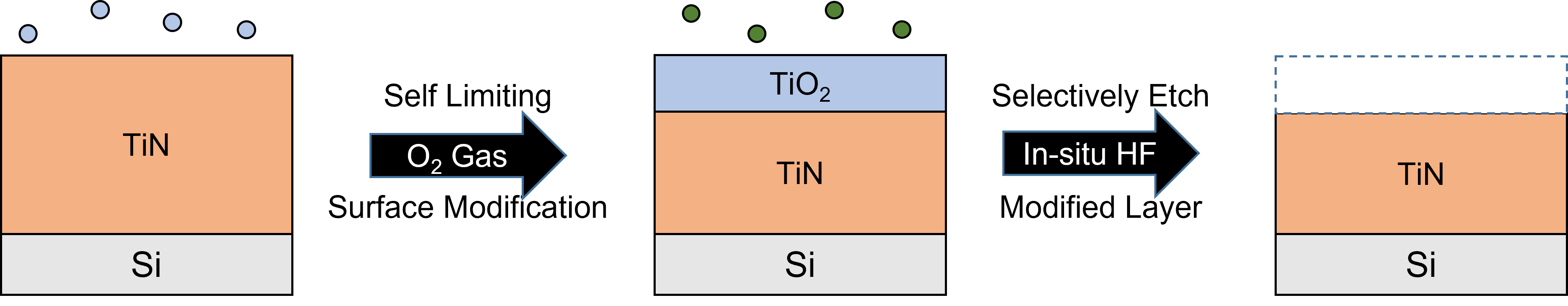}
    \caption{(a) Schematic of the TiN ALE process involving exposures to molecular oxygen to oxidize the surface (O\sub2, blue dots), followed by SF\sub{6}/H\sub{2} plasma (green dots) to produce volatile etch products.}
    \label{fig:TiNALEIntro}
\end{figure}
 
Here, we report the isotropic atomic layer etching of TiN using sequential exposures of O\sub{2} gas and SF\sub{6}/H\sub{2} plasma. The process is based on the selective etching of TiO\sub{2} over TiN for certain ratios of SF\sub{6}:H\sub{2}. The observed etch rates varied from 1.1 \Ang/cycle up to 3.2 \Ang/cycle for temperatures between \degC{150} and \degC{350} respectively, as measured using ex-situ ellipsometry. The etched surface was found to exhibit a $\sim 40$\% decrease in surface roughness. The superconducting transition temperature was unaffected by ALE beyond the expected change due to the decrease in film thickness, highlighting the low-damage nature of the process. Our findings indicate the potential of ALE in the processing of TiN for superconducting quantum electronics and microelectronics applications.

\section{Experiment}

The plasma-thermal ALE process of this work is illustrated in \Cref{fig:TiNALEIntro}. An exposure of molecular oxygen was used to oxidize the surface of TiN to TiO\sub{2}, followed by a purge. After, a mixture of SF\sub{6} and H\sub{2} gas was introduced into the chamber and ignited to form SF\sub{6}/H\sub{2} plasma. After this exposure, the reactor was again purged to complete the cycle. The use of SF\sub{6}/H\sub{2} plasma was motivated by noting that HF does not etch TiN, but fluorine radicals will spontaneously etch TiN \cite{Lee2017TiN, Pearton1991Aug}. Studies on SiN and Si etching using hydrogen and fluorine-containing plasma have shown that the plasma formed by the mixture yields different products at different plasma concentration ratios, including HF molecules at high hydrogen concentrations \cite{Pankratiev2020DecHF, Volynets2020HF, Jung2020Mar}. We therefore hypothesize that at high H\sub{2} concentrations, the SF\sub{6}/H\sub{2} plasma forms molecular HF which then selectively etches the TiO\sub{2} over the TiN, with minimal spontaneous etching from F radicals. The formation of HF in the SF\sub{6}/H\sub{2} plasma is referred to as ``in-situ HF" throughout the paper.

We investigated this approach to ALE of TiN using an Oxford Instruments FlexAL atomic layer deposition (ALD) system with an inductively-coupled plasma source, as described in Refs.~\cite{Coumou2012, vanHemmen2007May}. The substrate table temperature varied between \degC{150} to \degC{200}, as measured by the FlexAL substrate table thermometer. Prior to introducing the sample into the chamber for etching, the chamber walls and carrier wafer were conditioned by coating with 50 nm of Al\sub{2}O\sub{3} using 300 cycles of Al\sub{2}O\sub{3} ALD \cite{vanHemmen2007May}. Alumina was selected as it does not form volatile fluoride species on exposure to SF\sub{6} plasma. For TiN ALE, the sample was first exposed to 50 sccm O\sub{2} and 50 sccm Ar gas for 2 s at 100 mTorr pressure, followed by a 10 s purge. Next, a mixture of 20 sccm H\sub{2} and 4 sccm SF\sub{6} was stabilized at 100 mTorr for 5 s before striking the plasma at 100 W for 10 s. The excess reactants were purged for 10 s before repeating the cycle. The recipe resulted in a total time of $\sim 40$s per cycle. Before the sample was moved to the loadlock, the chamber was pumped down for 60 s. The sample was additionally held in the loadlock for two hours to cool down before exposure to air, so as to reduce oxygen diffusion into the sample.

The film thickness before and after etching was measured by ex-situ spectroscopic ellipsometry (J.A. Woolam M2000) at $60^{\circ}$ and $70^{\circ}$ from 370 nm to 1000 nm. Thickness was determined using 5 points on a 5 $\times$ 5 mm$^2$ square array. Subsequently, the data were fit using a Lorentz model to obtain the thickness of the samples \cite{Langereis2006Jul, Lee2017TiN}. Reported thickness values are the average of the 5 points. XPS analysis was performed using a Kratos Axis Ultra x-ray photoelectron spectrometer using a monochromatic Al K$\alpha$ source. Depth profiling was performed using an Ar ion beam with a 60 s interval for each cycle. The estimated milling depth was calculated based on initial and final film thickness measured by ex-situ ellipsometry and assuming a constant ion milling rate. The XPS data was analyzed in CASA-XPS from Casa Software Ltd. We adopt universal Tougaard background and sub-peak fitting routines from Refs.~\cite{Jaeger2012Nov, Maarouf2021Feb}.

The film surface topography was characterized using a Bruker Dimension Icon atomic force microscope (AFM) over a $0.25 \times 0.25\ \mu$m$^2$ area. The raw height maps collected on the AFM were processed by removing tilt via linear plane-fit. The surface roughness and power spectral density (PSD) were computed from the plane-fit height maps using procedures outlined in previous literature \cite{Gerritsen2022Dec, Jacobs2017Jan}. The PSD provides a quantitative measure of the lateral distance over which the surface profile varies in terms of spatial frequencies \cite{Elson1995Jan, Jacobs2017Jan}. The PSD was calculated by taking the absolute square of the normalized 1D-discrete Fourier transform of each row and column from the plane-fit AFM scan. The transformed data was then averaged to produce a single PSD curve. Reported roughness values and PSD curves were found to be consistent across 3 spots on each film.

Electrical resistivity measurements were performed on the Quantum Design DynaCool Physical Property Measurement System (PPMS). The TiN films were connected to the PPMS sample holder by four aluminum wires, wirebonded on the Westbond 7476D Wire Bonder. The film resistivity  $(\rho)$ was measured using a 4-point setup \cite{Shearrow2018Nov}. The resistivity was measured from 6 K to 1.7 K, and the data was used to calculate the superconducting critical temperature ($T_c$) of the films.

The samples consisted of 50 -- 60 nm thick TiN films on high resistivity Si (100) wafers ($>20\ \text{k}\Omega$cm, UniversityWafer) prepared using ALD with the same FlexAL system. The ALD process consisted of sequential half-cycles of exposure to tetrakis(dimethylamino)titanium (TDMAT) and nitrogen plasma with a 20 W DC bias at \degC{350}, similar to the procedure reported in Refs.~\cite{Shearrow2018Nov, Faraz2018Apr}. The resistance at 6 K and $T_c$ of a  60 nm thick ALD TiN film were measured to be $210\ \mu\Omega$cm and $3.22 \pm 0.06$ K, respectively; these values are comparable to those reported for other TiN films made using TDMAT \cite{Shearrow2018Nov, Musschoot2009Jan, Faraz2018Apr}. The chemical composition of the deposited films are described in \Cref{resistivityandTc}. The titania (TiO\sub{2}) films used for demonstrating etch selectivity in \Cref{sf6h2} were made by oxidizing TiN samples under an oxygen plasma for 5 minutes at \degC{300}, yielding a 5 nm thick TiO\sub{2} film on top of the TiN film. The thicknesses of the TiO\sub{2} films were measured using ex-situ ellipsometry.

\section{Results}

\begin{figure}
\centering
{\includegraphics[width = \textwidth]{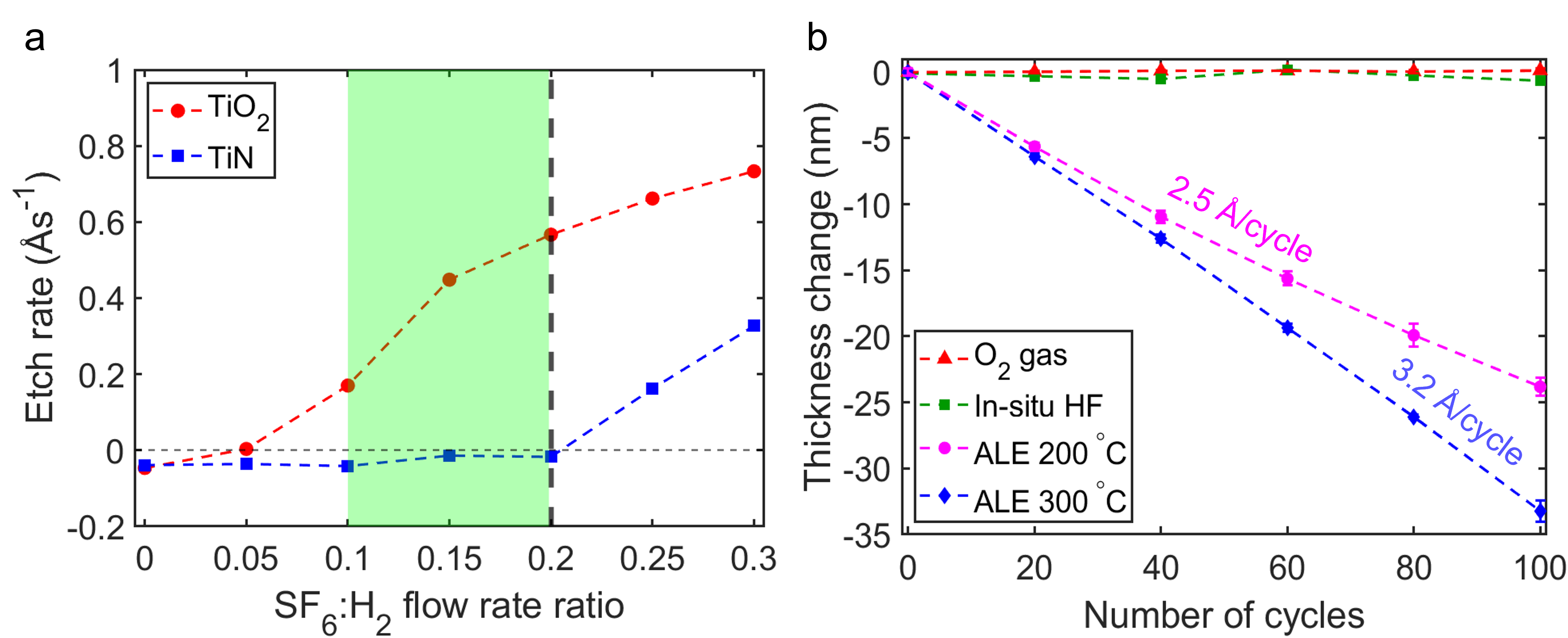}
    \phantomsubcaption\label{fig:insituHF}
    \phantomsubcaption\label{fig:thicknesscycle}}
    
\caption{(a) Etch rate of TiO\sub2 and TiN versus the SF\sub{6}:H\sub{2} flow rate ratio. The green shaded area represents the flow rate ratios for which selective etching of  TiO\sub2 over TiN was achieved. The black line at a ratio of 0.2 represents the ratio used in the ALE experiments. (b) TiN thickness change versus  number of cycles with exposure only to O\sub2 gas (red triangles), in-situ HF (green squares), full ALE process at \degC{200} (purple circles) and \degC{300} (blue diamonds). The dashed lines are guides to the eye.}

\label{fig:PseudoHF}
\end{figure}

\subsection{Selective etching with SF\sub{6}/H\sub{2} Plasma} \label{sf6h2}


We begin by examining the etch rate of TiO\sub{2} and TiN films for various SF\sub{6}:H\sub{2} flow rate ratios, $\eta$. \Cref{fig:insituHF} shows the etch rates of TiN and TiO\sub{2} versus $\eta$ at \degC{300}. For $\eta \lesssim 0.05$, negligible etching of either film is observed. Negative etch rates correspond to an increase in the thickness of the film, which we assume to be growth of non-volatile TiF\sub{3}. For $\eta \geq 0.1$, we observe spontaneous etching of TiO\sub{2}, with the etch rate  monotonically increasing with $\eta$. For TiN, we observe no etching for $\eta \leq 0.2$, but for $\eta \geq 0.25$ etching occurs. We attribute these observations to the formation of in-situ HF along with negligible fluorine radical concentration for $0.05 < \eta \leq 0.2$, similar to the results obtained in prior work \cite{Pankratiev2020DecHF, Volynets2020HF, Jung2020Mar}. For $\eta \geq 0.25$, the concentration of F radicals becomes sufficient to spontaneously etch the TiN, leading to increasing etch rates for both films. From our measurements, we find that $0.1 \leq \eta \leq 0.2$ achieves selective etching of TiO\sub2 over TiN. To obtain the highest etch selectivity of TiO\sub{2} over TiN, we select $\eta = 0.2$ for our experiments. This 1:5 ratio of SF\sub6:H\sub2 plasma is used throughout the rest of the paper.

\subsection{TiN ALE using O\sub2 and in-situ HF exposures}

\Cref{fig:thicknesscycle} shows the thickness change of TiN versus number of cycles for both half cycles, and for the full ALE recipe at \degC{200} and \degC{300}. For the half-cycles, the thickness change was measured after exposure to only molecular oxygen or in-situ HF. No etching was observed for either half-cycle, supporting the need for both steps. In contrast, we observe a decrease in the thickness with increasing number of cycles when using both steps. The etch rate is calculated by dividing the total thickness change by the number of cycles, giving values of  $2.5 \pm 0.16$ \Ang/cycle at \degC{200} and $3.2 \pm 0.10$ \Ang/cycle at \degC{300}.

\begin{figure}
\centering
{\includegraphics[width = \textwidth]{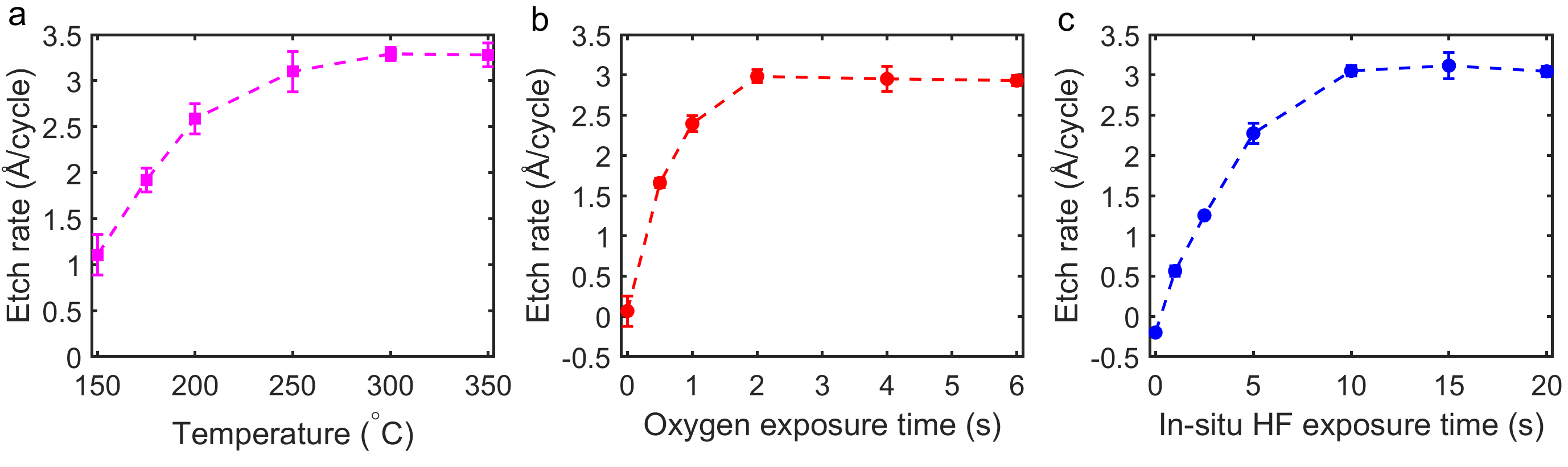}
    \phantomsubcaption\label{fig:Temp}
    \phantomsubcaption\label{fig:O2sat}
    \phantomsubcaption\label{fig:HFsat}}

\caption{(a) TiN ALE etch per cycle (EPC) versus substrate table temperature. (b) EPC versus O\sub2 gas exposure time with in-situ HF exposure time fixed at 10 s. (c) EPC versus in-situ HF time with O\sub2 exposure time fixed at 2 s. The etch rates are observed to saturate with exposure time, demonstrating the self-limiting nature of the ALE process.}

\label{fig:Saturation}
\end{figure}

We further examine the effect of temperature on the etch rate. \Cref{fig:Temp} shows the etch per cycle (EPC) versus table temperature ranging from \degC{150} and \degC{350}. The etch rates are calculated over a 100 cycles using a linear fit. We find that the etch rate increases from 1.1 \Ang/cycle at \degC{150} to 3.2 \Ang/cycle at \degC{300}. The increase in EPC with temperature is similar to what has been observed in previous thermal ALE studies of various materials \cite{Lee2017TiN, Shim2022, George2020Jun, Chittock2020Oct}. We also observe a constant etch rate from \degC{300} to \degC{350}, similar to what is reported in Figure 7 of Ref.~\cite{Lee2017TiN}.

We also explored the self-limiting nature of the recipe by measuring the saturation curves of each half-cycle. For each saturation curve, the purge times and one half-cycle time are fixed while the other is varied. The etch rates reported are calculated based on the thickness change over 50 cycles at \degC{300}. In \Cref{fig:O2sat}, the in-situ HF step is fixed at 10 s, while the etch rate is measured versus the oxygen exposure time. The etch rate is observed to saturate to $\sim 3$ \Ang/cycle above 2 s, which is consistent with the self-limiting nature of the oxidation step. In \Cref{fig:HFsat}, the oxidation step is fixed at 2 s, while the etch rate is measured versus in-situ HF exposure time. The etch rate saturates to $\sim 3$ \Ang/cycle above 10 s, which is consistent with the selectivity of the in-situ HF to etch TiO\sub{2} and terminate on the TiN.

\subsection{Characterization of film composition}

\begin{figure}
\centering
{\includegraphics[width = \textwidth]{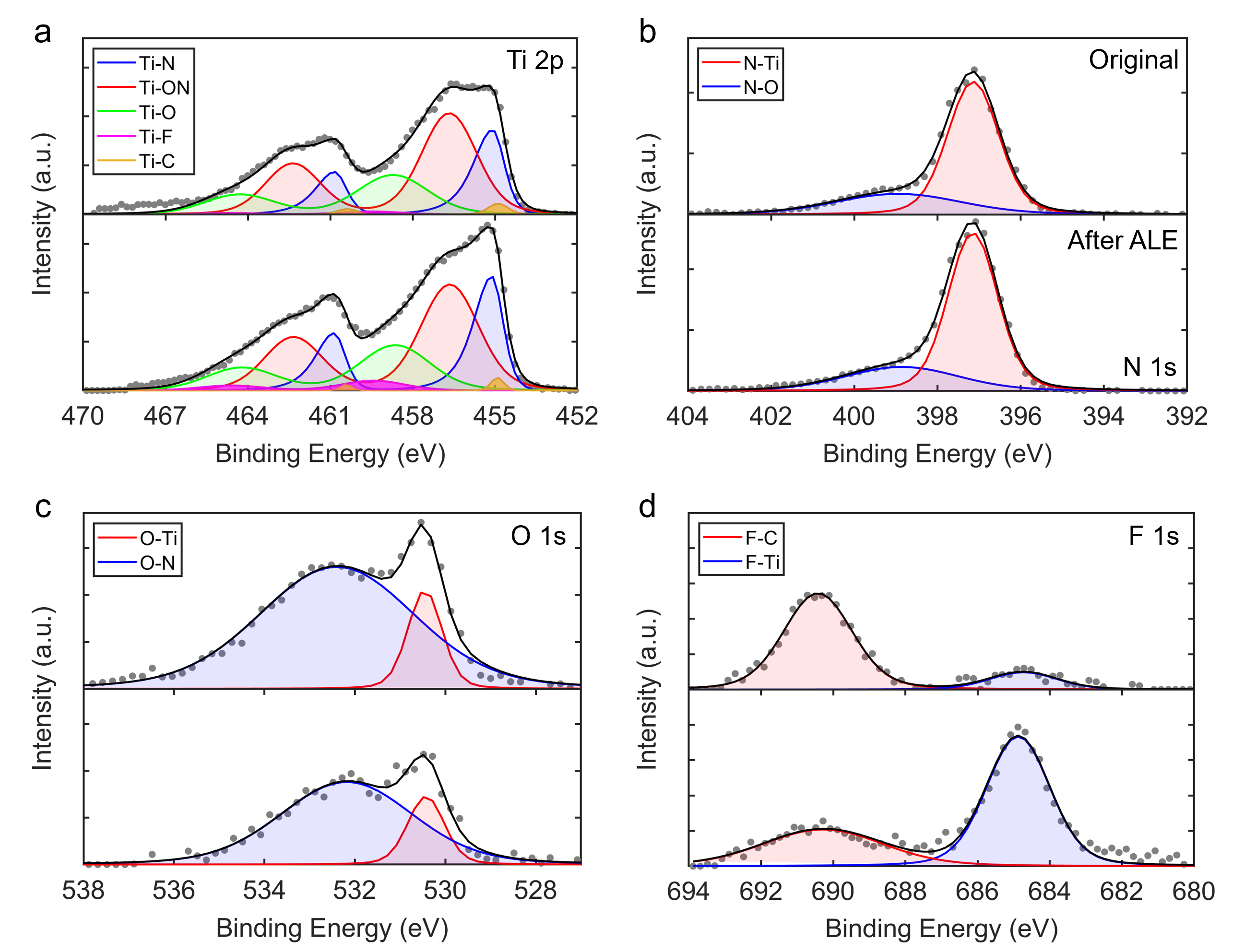}
    \phantomsubcaption\label{fig:Ti2p}
    \phantomsubcaption\label{fig:N1s}
    \phantomsubcaption\label{fig:O1s}
    \phantomsubcaption\label{fig:F1s}
}
\caption{Surface XPS spectra showing (a) Ti2p, (b) N1s, (c) O1s and (d) F1s spectra. The spectra is shown for (top) original and (bottom) etched TiN films. The measured (gray dots) and fit spectra (black lines) intensity are reported in arbitrary units (a.u.) against the binding energy on the x-axis.}
\label{fig:XPSSurface}
\end{figure}

We next characterize the chemical composition of the TiN films before and after ALE using XPS. In \Cref{fig:XPSSurface}, we show the core levels of Ti2p, N1s, O1s, C1s and F1s. For the Ti2p XPS spectra in \Cref{fig:Ti2p}, we observe five components. Each component is a doublet consisting of a 2p\sub{3/2} and a 2p\sub{1/2} subpeak. We observe subpeaks corresponding to Ti-C (454.9 eV and 460.4 eV) \cite{Santerre1999Jun, Naslund2020Dec, Luthin2001}, Ti-N (455.1 eV and 460.8 eV) \cite{Kuznetsov1992Mar, Saha1992Oct, Chan2009Jul}, Ti-ON (456.5 eV and 462.3 eV) \cite{Kuznetsov1992Mar, Saha1992Oct, Chan2009Jul}, Ti-O (458.5 eV and 464.2 eV) \cite{Kuznetsov1992Mar, Saha1992Oct, Chan2009Jul}, and Ti-F (459.4 eV and 465.6 eV) \cite{Mousty1987Apr, Natu2021Apr}. In \Cref{fig:N1s}, we report the N1s spectra with two subpeaks at 397.1 eV and 398.9 eV, belonging to N-Ti and N-O bonds, respectively \cite{Kuznetsov1992Mar, Saha1992Oct, Chan2009Jul}. In \Cref{fig:O1s}, we report the O1s spectra with two subpeaks at 530.4 eV and 532.2 eV, corresponding to O-Ti and O-N bonds, respectively \cite{Kuznetsov1992Mar, Saha1992Oct, Chan2009Jul}. In \Cref{fig:F1s}, we report the F1s spectra with two subpeaks at 684.9 eV and 690.3 eV, corresponding to F-Ti and F-C bonds, respectively \cite{Mousty1987Apr, Beamson1993Jan, Natu2021Apr}. 

We observe that the Ti2p spectra is dominated by oxides and oxynitrides, consistent with the presence of a native oxide on TiN 
\cite{Musschoot2009Jan, Saha1992Oct}. After ALE (bottom panels of \Cref{fig:Ti2p,fig:N1s,fig:O1s}), an increase in the magnitude of the Ti-N and N-Ti peaks is observed along with an overall decrease in the O1s peak magnitude. The decreased O1s signal implies a reduced native oxide concentration after ALE, as has been observed in other works \cite{Wang2023May, Hennessy2017Jul, Metzler2017Jun}. The F1s spectra for the original sample may be attributed to contamination from using the same chamber for deposition and etching, which is consistent with the reduced magnitude of the F1s peak in the original sample compared to that in the ALE-treated sample  (bottom panel of \Cref{fig:F1s}).

\begin{figure}
\centering
{\includegraphics[width = \textwidth]{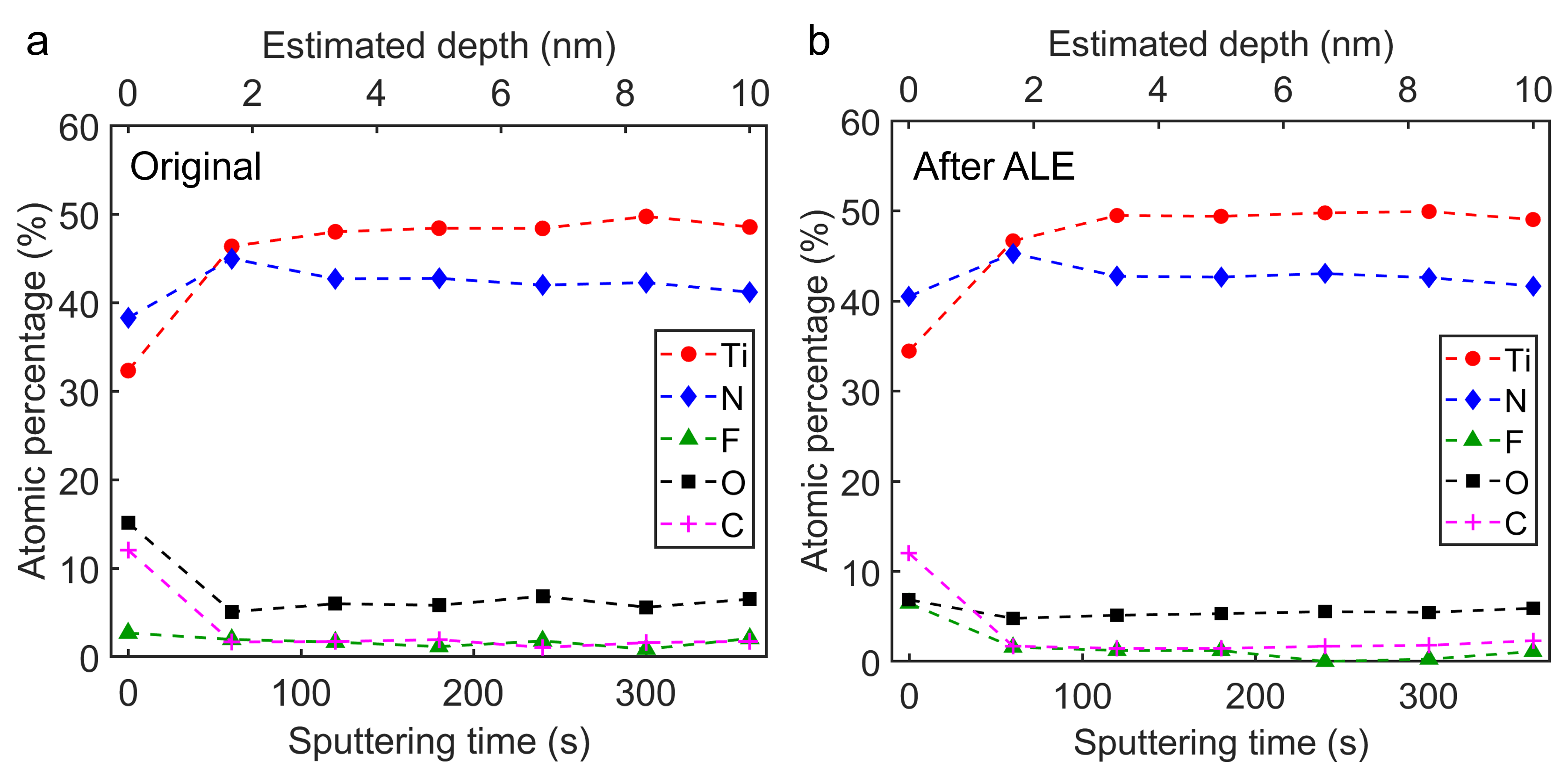}
    \phantomsubcaption\label{fig:xpsbefore}
    \phantomsubcaption\label{fig:xpsafter}
}

\caption{Atomic concentrations of Ti, N, O, F, and C versus Ar milling time and estimated depth, for (a) original and (b) ALE-treated TiN thin film.}
\label{fig:XPSDepth}
\end{figure}

We also performed depth-profiling XPS to determine the atomic concentrations on the surface and bulk. In \Cref{fig:XPSDepth}, we show the atomic concentrations of Ti, N, F, C and O as a function of sputtering time and estimated depth in the original and ALE-treated films. In the original sample (\Cref{fig:xpsbefore}), the atomic concentrations on the surface are 31.9\% (Ti), 37.6\% (N), 16.1\% (O), 12.0\% (C), and 2.4\% (F). After 120 s Ar milling $(\sim 3.5\ \text{nm})$, the atomic concentrations plateau to their bulk values of 48.6\% (Ti), 42.3\% (N), 6.1\% (O), 1.9\% (C), and 1.1\% (F). The carbon and oxygen levels are consistent with other reported ALD TiN films made using TDMAT \cite{Musschoot2009Jan, Fillot2005Dec, Elam2003Jul}. For the ALE-treated sample (\Cref{fig:xpsafter}), the atomic concentrations on the surface are 34.2\% (Ti), 39.5\% (N), 7.9\% (O), 11.9\% (C), and 6.5\% (F). After 120 s Ar milling $(\sim 3.5\ \text{nm})$, the atomic concentrations plateau to their bulk values of 49.0\% (Ti), 42.2\% (N), 5.9\% (O), 1.8\% (C), and 1.1\% (F). We observe a $\sim 49\%$ decrease in the surface oxygen concentration in the ALE-treated film. An increase in the surface fluorine concentration of the ALE-treated film is also observed, consistent with other works involving the interactions of fluorine-containing plasma with dielectric films \cite{Fischer2017Mar, Wang2023May}. The atomic concentrations in the bulk of the ALE-treated film are within 95\% of the values in the original film. Therefore, we conclude that the effect of ALE is confined to a few nanometers of the surface, with negligible effect on the bulk chemical composition.

\subsection{Surface roughness characterization}

\begin{figure}
\centering
{\includegraphics[width = 1\textwidth]{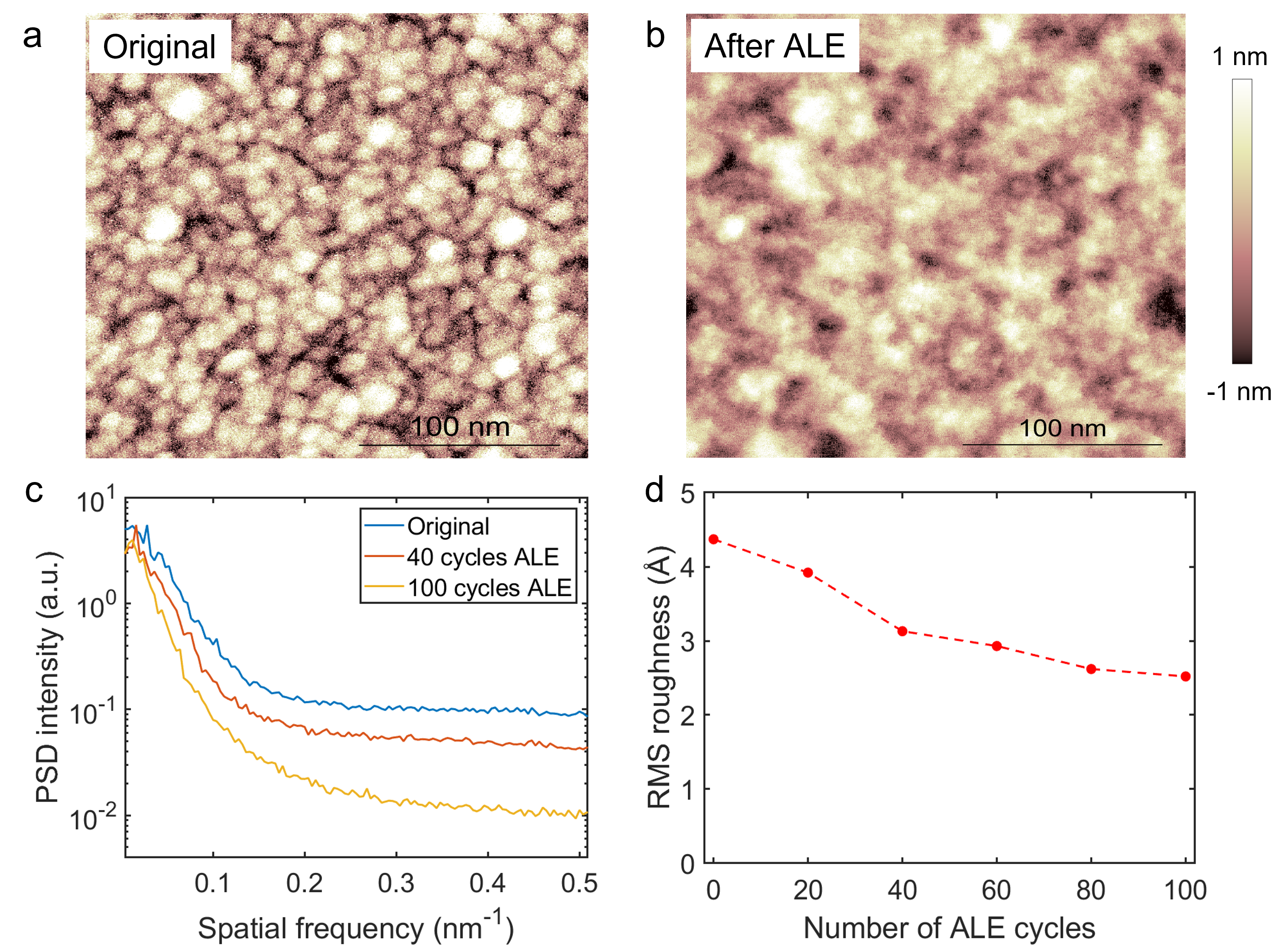}
    \phantomsubcaption\label{fig:AFMALD}
    \phantomsubcaption\label{fig:AFMALE}
    \phantomsubcaption\label{fig:PSD}
    \phantomsubcaption\label{fig:Roughness}}

\caption{AFM scan showing height-maps of original ALD sample (a) and after 100 ALE cycles (b). (c) Progression of the height-map PSD with increasing ALE cycles, showing decrease across all spatial frequencies. (d) RMS roughness computed from AFM height-map against number of ALE cycles.}
\label{fig:AFMHeight}

\end{figure}

We characterized the roughness of the TiN films before and after ALE using AFM. \Cref{fig:AFMALD} shows the plane-fit height map of the film as deposited using ALD. \Cref{fig:AFMALE} shows the plane-fit height map after 100 cycles of ALE at \degC{300}. \Cref{fig:PSD} shows the PSD curves for the original film, after 40 ALE cycles and after 100 ALE cycles at \degC{300}. We observe a decrease in the PSD intensity across all length scales as the number of ALE cycles is increased, indicating that features with length scales from $\sim 2 - 20$ nm are smoothed by the ALE process. In \Cref{fig:Roughness}, the RMS roughness is plotted versus the number of ALE cycles at \degC{300}. We observe a monotonic decrease in RMS roughness from 4.4 \Ang\ to 2.5 \Ang~after 100 cycles. This 43\% reduction in roughness was observed across 3 different positions on the sample.

\subsection{Electrical and superconducting properties} \label{resistivityandTc}

\begin{figure}
    \centering
    \includegraphics[width = 0.6\textwidth]{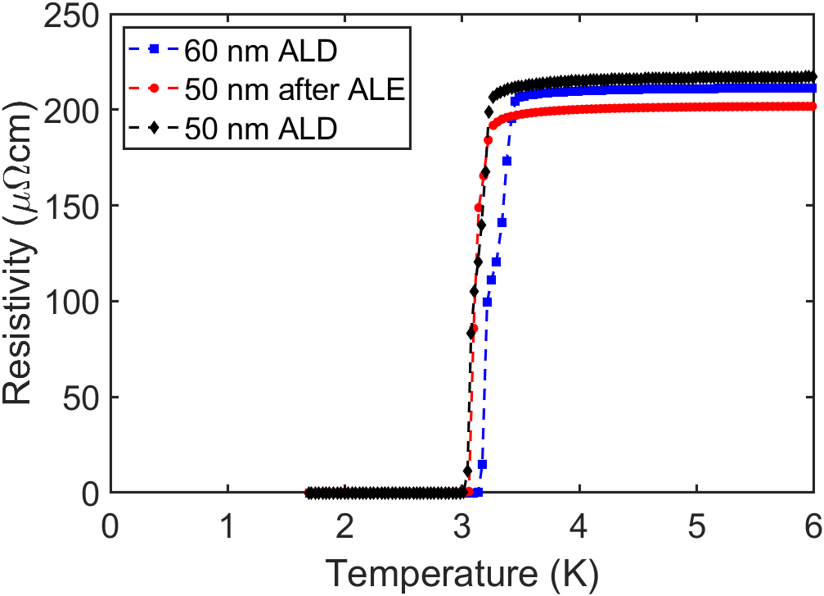}
    \caption{Resistivity versus temperature for original 60 nm TiN film (blue), ALE-treated film of 50 nm thickness (red) and a 50 nm ALD TiN film (black) for comparison. The difference in $T_c$ between the 50 nm ALD  film and ALE-treated 50 nm film is negligible. The dashed lines are guides to the eye.}
    \label{fig:PPMS}
\end{figure}

We investigated the effect of ALE on the electrical and superconducting properties of the TiN films by measuring their resistivity from 6 K to 1.7 K. A 60 nm TiN film was deposited using ALD, which was etched to 50 nm using ALE. Another 50 nm TiN film was prepared using ALD to compare to the ALE-treated 50 nm film. The measured resistivity versus temperature for the three films is shown in \Cref{fig:PPMS}. The resistivity at 6 K of the 60 nm ALD film is found to be $222\ \mu\Omega$cm, with a superconducting critical temperature $T_c = 3.22 \pm 0.06$ K. The resistivity of the TiN film is consistent with those previously reported for ALD TiN films \cite{Shearrow2018Nov, Musschoot2009Jan}, and the $T_c$ reported is similar to the $T_c$ of other TiN films grown with TDMAT \cite{Shearrow2018Nov, Proslier2011Oct}. After 40 cycles of ALE at \degC{200}, the TiN thickness decreased to 50 nm, with a resistivity of $201\ \mu\Omega$cm at 6 K and $T_c = 3.13 \pm 0.04$ K. For comparison, the 50 nm ALD film had a resistivity of $227\ \mu\Omega$cm at 6 K, and $T_c= 3.11 \pm 0.05$ K. We therefore find that the change in $T_c$ of the TiN film after ALE is consistent with that expected with a decrease of 10 nm in thickness, without any additional decrease due to process-induced damage. This observation highlights the improved quality of the processed films compared to those obtained from processing methods which lack atomic control \cite{Zheng2022Mar}. The reduced 6 K resistivity of the ALE-treated film is thought to arise due to the removal of the native oxide. This result warrants further investigation and is a topic of future study.

\section{Discussion}

We now discuss the characteristics of our plasma-thermal TiN ALE process in context with isotropic thermal ALE processes for TiN and related materials. Thermal ALE of TiN has been reported using molecular O\sub{3} or H\sub{2}O\sub{2} and HF vapor \cite{Lee2017TiN}, and O\sub{2} plasma and CF\sub{4} plasma \cite{Shim2022}. The first process leads to an etch per cycle (EPC) of 0.20 \Ang/cycle at \degC{200}, achieving atomic-scale control of etching. However, the recipe requires the use of HF vapor which incurs practical complications. The second process based on O\sub{2} plasma and CF\sub{4} plasma achieves an EPC of 17.1 \Ang/cycle at \degC{200}, which is a larger EPC than is desired for manipulating the surface region of the films. The second process also requires an additional heating step, which can lead to impractical process times on conventional tools. The present recipe achieves an EPC of 2.6 \Ang/cycle at \degC{200} and 3.1 \Ang/cycle at \degC{250}, providing etch rates between the previous reported recipes. The present recipe also avoids the use of HF, requiring only an SF\sub{6}/H\sub{2} plasma that also yields etching selectivity of TiO\sub{2} over TiN.

Our isotropic plasma-thermal ALE may find potential applications in the fabrication of TiN-based superconducting microresonators for microwave kinetic inductance detectors and qubits, where the native oxide hosts parasitic TLS that presently limit the device performance. Based on our XPS and resistivity measurements, ALE-treated films have a reduced oxygen concentration while maintaining unaltered bulk chemistry and electrical properties. These properties make ALE promising for reducing the number of TLS in the metal-air interface and thereby improving the quality factor of superconducting microresonators. The smoothing effect and isotropic Angstrom-scale EPC of the present ALE recipe is also relevant for fabricating TiN-based nanoscale metal gate electrodes in CMOS devices and various transistor designs, where the metal layers are required to have thickness on the order of $\sim 10$ nm with uniformity $\lesssim 4\%$ \cite{Zhao2019Jun, Matsukawa2015Jun}. The ALD system in our work (Oxford Instruments, FlexAL) has demonstrated high uniformity on 200 mm diameter substrates \cite{vanHemmen2007May}, and therefore our process has the potential to extend to wafer-scale applications.

\section{Conclusion}

We have reported an isotropic plasma-thermal atomic layer etching process for TiN using sequential exposures of molecular oxygen and SF\sub{6}/H\sub{2} plasma. The SF\sub{6}/H\sub{2} plasma selectively etches TiO\sub{2} over TiN for SF\sub{6}:H\sub{2} flow rate ratios between 0.1 and 0.2. The etch rate varies from 1.1 \Ang/cycle at \degC{150} to 3.2 \Ang/cycle at \degC{350}. We observe a smoothing effect from ALE, corresponding to a $\sim 43\%$ reduction in RMS roughness after 100 cycles. The surface oxygen concentration is reduced by $\sim 49\%$ after 100 cycles of ALE, indicating a decrease in the volume of surface oxide. We also find that ALE does not induce any change in $T_c$ beyond that expected from the decrease in film thickness, highlighting the low-damage nature of the process. We anticipate that the ability to engineer the surface of TiN films on the Angstrom-scale using isotropic ALE will facilitate applications of TiN in superconducting resonators and microelectronics.

\section{Acknowledgements}

This work was supported by NSF under Award \#2234390. The authors thank Nicholas Chittock (Eindhoven University of Technology) for useful discussions, and Phillipe Pearson (California Institute of Technology) for assistance with the wirebonder. We gratefully acknowledge the critical support and infrastructure provided for this work by The Kavli Nanoscience Institute and the Molecular Materials Research Center of the Beckman Institute at the California Institute of Technology.

\newpage
\clearpage

\bibliography{ref}

\end{document}